\documentclass[%
reprint,
superscriptaddress,
nofootinbib,
amsmath,
amssymb,
aps,
floatfix,
showkeys,
]{revtex4-2}

\usepackage[normalem]{ulem}

\usepackage{fancyhdr}
\usepackage{graphicx,amsfonts,amssymb,amsbsy}
\usepackage{amsmath,amsthm,latexsym}
\usepackage[utf8]{inputenc}  

\usepackage{natbib}
\usepackage[colorlinks]{hyperref}
\hypersetup{linkcolor=magenta,citecolor=blue}

\begin{document}


\title{Hybrid stars with reactive interfaces: analysis within the  Nambu-Jona-Lasinio model}

\author{C. H. Lenzi}
\email{chlenzi@ita.br}
\affiliation{Department of Physics, Instituto Tecnológico de Aeronáutica, DCTA, CEP 12228-900, São José dos Campos, SP, Brazil}

\author{G. Lugones}
\email{german.lugones@ufabc.edu.br}
\affiliation{Universidade Federal do ABC, Centro de Ciências Naturais e Humanas, Avenida dos Estados 5001- Bangú, CEP 09210-580, Santo André, SP, Brazil.}

\author{C. Vasquez}
\email{cesar.vasquez@uemasul.edu.br}
\affiliation{Universidade Estadual da Região Tocantina do Maranhão, Centro de Ciências Exatas, Naturais e Tecnológicas,  Imperatriz, CEP 65901-480, MA, Brazil}
\affiliation{Universidade Federal do Maranhão, Departamento de Física-CCET, Campus Universitário do Bacanga, São Luís, MA, CEP 65080-805, Brazil}

\begin{abstract}
It has been shown recently that quark-hadron conversions at the interface of a hybrid star may have a key role on the dynamic stability of the compact object. In this work we perform a systematic study of hybrid stars with reactive interfaces using a model-agnostic piecewise-polytropic hadronic equation of state and the Nambu-Jona-Lasinio model for three-flavor quark matter. For the hadronic phase we use a soft, an intermediate and a stiff parametrization that match at $1.1 n_0$  {with predictions} based on chiral effective field theory (cEFT) interactions. 
In the NJL Lagrangian we include scalar, vector and 't Hooft interactions. The vector coupling constant $g_{v}$ is treated as a free parameter. We also consider that there is a split between the deconfinement and the chiral phase transitions which is controlled by changing the conventional value of the vacuum pressure $-\Omega_{0}$ in the NJL thermodynamic potential by $-\left(\Omega_{0}+\delta \Omega_{0}\right)$, being $\delta \Omega_{0}$ a free parameter. 
We analyze the mass-radius ($M$-$R$) relation in the case of rapid ($\tau \ll 1 \, \mathrm{ms}$) and slow ($\tau \gg 1 \, \mathrm{ms}$) conversions, being $\tau$ the reaction timescale. In the case of slow interface reactions we find $M$-$R$ curves with a cusp at the maximum mass point where a pure hadronic branch and a slow-stable hybrid star (SSHS) branch coincide. We find that the length of the slow-stable branch grows with the increase of the transition density and the energy density jump at the hadron-quark interface. We calculate the tidal deformabilities of SSHSs and analyse them in the light of the  GW170817 event. 
\end{abstract}

\maketitle

\section{Introduction}

A reliable description of the equation of state (EOS) of dense matter above $\sim 1 - 2 \,n_0$ (being $n_{0}=0.16 \, \mathrm{fm}^{-3}$ the nuclear saturation density) still faces many complications that make difficult to infer the internal composition of neutron stars (NSs). One possible candidate among the wide variety of internal compositions proposed in the literature are the so-called hybrid stars composed of an inner quark matter core and a hadron matter mantle. However, not only it is still unclear whether a hadron-quark phase transition would occur at typical densities of the NS interior, but it is also unknown whether the transition is of first order or a crossover \cite{Fukushima:2013rx}. 

Assuming that  the transition is of first order in the high-density low-temperature regime of the QCD phase diagram, it is also unclear whether quarks and hadrons would be separated by a sharp discontinuity or by a mixed phase where both phases coexist along a wide density region forming globally charge-neutral geometrical structures. Such mixed phases are energetically preferred if finite size effects are not too large. In practice, the most important ingredient is the quark matter surface tension $\sigma$.
Mixed phases are favored if $\sigma$ is smaller than a critical value $\sigma_\mathrm{crit}$ of the order of tens of $\mathrm{MeV} / \mathrm{fm}^{2}$; but if $\sigma > \sigma_\mathrm{crit}$, the mixed phase is unstable and a sharp interface would be formed \citep{Wu:2018zoe,Maslov:2018ghi}. 

Unfortunately, theoretical values of $\sigma$ span a wide range depending on the EOS and on the calculation method:  some authors  obtain  $\sigma < \sigma_\mathrm{crit}$ \citep{Garcia:2013eaa,Lugones:2016ytl,Lugones:2018qgu,Fraga:2018cvr,Gao:2016hks} but  very large $\sigma$ favouring a sharp interface is obtained using the multiple reflection expansion method with the Nambu-Jona-Lasinio EOS  \citep{Lugones:2013ema} and the MIT bag EOS with vector interactions \citep{Lugones:2021tee}.
In this work, we adopt as a working hypothesis that the  interface is sharp. 

Hybrid stars with sharp discontinuities have been widely studied in recent years (see e.g. Refs. \cite{Alford:2004pf, Negreiros:2010tf, Alford:2013aca, Lenzi:2012xz, Burgio:2015zka, Kaltenborn:2017hus, Paschalidis:2017qmb, Li:2021sxb, Li:2019fqe, Cierniak:2020eyh, Otto:2019zjy} and references therein). However, no attention was given in these works to the behavior of the quark-hadron interface when the star is dynamically perturbed. Such an analysis is of fundamental importance because in the presence of reactive discontinuities, perturbations may induce the phase conversion of fluid elements around the interface \cite{Pereira:2017rmp}. 
%
%
 {Unfortunately, the actual conversion timescale at the interface remains uncertain. Since phase transitions are highly collective and nonlinear phenomena, the transition timescale cannot be solely attributed to independent particle interactions. Therefore, despite the typical interaction timescales for strong and weak forces being $\sim 10^{-23} \mathrm{s}$ and $\sim 10^{-8} \mathrm{s}$, respectively, the conversion timescale in a perturbed hybrid star is not necessarily fast. 
Nucleation, identified as the primary driving force behind first-order phase transitions in various systems (for example, see Ref. \cite{Slezov2009}), is a potential mechanism for first-order hadron-quark transitions at high densities \citep{Olesen:1993ek, Lugones:1997gg, Iida:1998pi, Bombaci:2004mt}. In NSs, this process faces a significant activation barrier, as direct conversion of hadronic matter to quark matter in equilibrium under weak interactions is typically a high-order weak process that is strongly suppressed. Therefore, nucleation is expected to proceed through an intermediate state that is not easily accessible near the interface (see Figs. $1$ and $2$ of Ref. \cite{Lugones:2015bya}). The quantum and thermal nucleation timescales have been calculated and found to be larger than the age of the universe for temperatures below a few MeV. Note that this timescale drops significantly at higher temperatures \cite{Bombaci:2016xuj}. Similarly, for the reverse reaction, a high-order weak interaction process is needed within quark matter to produce beta-equilibrated hadronic matter with a lower free energy, making the process highly unlikely. Therefore, while the conversion timescale is uncertain, there is compelling evidence to suggest that it could be slow.
Another conversion mechanism based on strangeness diffusion in a simple tube model has been proposed \cite{Alford:2014jha, Olinto:1986je}, which seems powerful enough to saturate $r$-modes at very low amplitude, of order $10^{-10}$. If a similar mechanism operates for radial oscillations, interface conversion could be rapid.  
Despite the aforementioned uncertainties, the stellar response in scenarios of extremely rapid or extremely slow conversions is insensitive to the kinetic details of the conversion process, allowing robust studies in these limits \cite{Pereira:2017rmp,Lugones:2021bkm}. Since the typical oscillation period of compact objects is in the ballpark of $1 ~ \mathrm{ms}$, phase conversions can be said to be rapid when $\tau_{\mathrm{conv}} \ll 1 ~ \mathrm{ms}$ and slow when $\tau_{\mathrm{conv}} \gg 1 ~ \mathrm{ms}$.
}

The first study of the role of hadron-quark conversions  on the dynamic stability of hybrid stars was carried out in Ref. \cite{Pereira:2017rmp} and showed that changes of stellar stability do not occur at maxima or minima in the stellar mass $M$ versus central density $\epsilon_{c}$ diagram if phase conversions are slow. As a consequence, some stellar configurations with $\partial M / \partial \epsilon_{c}<0$ may be dynamically stable.
Later studies analyzed various aspects of hybrid stars with sharp interfaces and slow reactions (the so called slow-stable hybrid stars, SSHSs) using several phenomenological models for quarks and hadrons \cite{Parisi:2020qfs, DiClemente2020, Tonetto:2020bie, Lugones:2021bkm, Mariani:2019vve, Rodriguez:2020fhf, Ranea:2022bou,Rau:2022ofy}. However, SSHSs have not been studied yet within the Nambu-Jona-Lasinio (NJL) model for quark matter, which is the focus of the present work.
 {Indeed, previous studies of SSHS have mostly relied on simple phenomenological models, such as the MIT bag model (see \cite{Parisi:2020qfs}), or model-agnostic EOSs like the constant speed of sound EOS (see \cite{Lugones:2021bkm}). Although these analyses provide a general understanding of SSHS properties, their predictions may be incompatible with models that incorporate a more detailed description of the underlying microphysics, potentially leading to extreme scenarios.
By contrast, the Nambu-Jona-Lasinio (NJL) model is a well-established phenomenological model of quark matter that incorporates important QCD ingredients, such as the chiral phase transition, as well as a variety of interaction terms that describe the dynamics of quarks and their interactions with each other. It can accurately predict data on meson masses and decay constants and plays a significant role in clarifying other characteristics of quark matter, such as its phase structure and related properties.
Through the utilization of the NJL model, our current study of SSHS aims to enhance the characterization of these objects by narrowing down the range of possible outcomes that result from the use of generic EOSs.}
To this end, we will use the NJL Lagrangian of Ref. \cite{Lenzi:2012xz} which includes scalar, vector and ’t Hooft interactions and takes the vector coupling constant $g_{v}$ as a free parameter. We will also consider that there is an arbitrary split between the deconfinement and the chiral phase transitions. This split can be tuned by a redefinition of the ``bag'' constant parameter $\Omega_0$ in the NJL thermodynamic potential, which can be implemented by changing $\Omega_0$ by $\Omega_0 + \delta \Omega_0$ and taking $\delta \Omega_0$ as a free parameter.
In  Ref. \cite{Lenzi:2012xz} we found that, as  $\delta \Omega_0$ and/or $g_v$  are increased, hybrid stars have a larger maximum mass but they are stable within a smaller range of central densities. For large enough $\delta \Omega_0$ and $g_v$, stable hybrid configurations are not possible at all. 
In this work we will revisit the analysis of Ref. \cite{Lenzi:2012xz}, but taking into account the possibility of a reactive hadron-quark interface and putting special emphasis in SSHSs.

The paper is organized as follows. In Section \ref{sec:2} we describe the EOSs that will be used in our calculations. In Section \ref{sec:MR} we analyse the mass-radius relation and in Section \ref{sec:tidal} the tidal deformability of hybrid stars with reactive interfaces, showing that they are in agreement with current observations. In Section \ref{sec:4} we summarize our results and explore some of their consequences.

\begin{table}[tb]
\centering
\begin{tabular}{cccc}
\toprule
\quad Model \quad & \quad $\Gamma_1$ \quad &  \quad $\Gamma_2$  \quad & \quad $\Gamma_3$  \quad \\
\toprule
Soft           & 2.752 & 4.5 & 3.5  \\ 
\hline
Intermediate   & 2.758 & 6.5 & 3.2  \\ 
\hline
Stiff          & 2.764 & 8.5 & 3.2  \\
\toprule
\end{tabular}
\caption{Parameters of the model-agnostic GPP fit \cite{OBoyle-etal-2020} used for the hadronic part of the EOS with baryon number density above $0.3\, n_0$. In all cases, we adopted $\log_{10}K_1 = -27.22$, $\log_{10} \rho_0 = 13.902$,  $\log_{10} \rho_1 = 14.45$ and  $\log_{10} \rho_2 = 14.58$.}
\label{table:hadronic}
\end{table}

\begin{table*}[ht]
\footnotesize
\setlength{\tabcolsep}{3pt} 
\renewcommand{\arraystretch}{0.8} 
\begin{tabular}{|ccc|cccccc|cccccc|cccccc|} 
\toprule
\multicolumn{3}{|c|}{} & \multicolumn{6}{c|}{Soft} & \multicolumn{6}{c|}{Intermediate} & \multicolumn{6}{c|}{Stiff} \\  
Set  &  ${g_v}/{g_s}$ & $\delta\Omega_0$ & $\Delta \varepsilon$   & $P_t$ & $M_{\mathrm{T}}$ & $R_{\mathrm{T}}$ & 
$M_{\mathrm{max}}$   & $R_{\mathrm{max}}$ & $\Delta \varepsilon$ & $P_t$ & $M_{\mathrm{T}}$  & $R_{\mathrm{T}}$ & 
$M_{\mathrm{max}}$   & $R_{\mathrm{max}}$ & $\Delta \varepsilon$   & $P_t$ & $M_{\mathrm{T}}$ & $R_{\mathrm{T}}$ & 
$M_{\mathrm{max}}$   & $R_{\mathrm{max}}$ \\
\toprule
1 &   0.0  & -18.3 & 71   & 103 & 1.79 & 11.1 & $1.80^*$  & 11.3 & $-$  & $-$  &  $-$ &  $-$ & $-$      &  $-$ &  $-$ & $-$  & $-$  &  $-$ & $-$      &  $-$ \\
2 &   0.0  & -9.1  & 200  & 154 & 1.78 & 10.9 & $1.88^*$  & 11.6 & 127  & 52   & 1.75 & 11.5 & $1.77^*$ & 11.7 & 155  & 32   & 1.72 & 11.5 & $1.72^*$ &  11.7 \\
3 &   0.0  & 0.0   & 383  & 189 & 1.66 & 10.3 & 1.97      & 11.6 & 197  & 77   & 1.78 & 11.7 & $1.85^*$ & 12.5 & 220  & 47   & 1.73 & 11.8 & $1.78^*$ &  14.1 \\
4 &   0.0  & 9.1   & 531  & 202 & 1.57 & 9.9  & 2.00      & 11.6 & 265  & 97   & 1.77 & 11.6 & 1.99     & 11.7 & 287  & 59   & 1.75 & 12.0 & 2.01     &  14.3 \\
5 &   0.0  & 18.3  & 654  & 216 & 1.52 & 9.7  & 2.03      & 11.6 & 326  & 115  & 1.71 & 11.3 & 2.10     & 13.2 & 335  & 70   & 1.73 & 11.9 & 2.18     &  14.5 \\
6 &   0.0  & 91.4  & 1377 & 288 & 1.46 & 8.6  & 2.15      & 11.5 & 1163 & 191  & 1.47 & 10.3 & 2.38     & 13.0 & 986  & 131  & 1.50 & 11.0 & 2.70     &  14.6 \\
7 &   0.0  & 137.1 & 1733 & 320 & 1.47 & 8.8  & 2.18      & 11.4 & 1514 & 219  & 1.48 & 10.4 & 2.44     & 12.9 & 1361 & 152  & 1.55 & 11.1 & 2.80     &  14.6 \\
8 &   0.0  & 182.7 & 2012 & 349 & 1.47 & 8.8  & 2.21      & 11.3 & 1831 & 242  & 1.50 & 10.5 & 2.48     & 12.9 & 1727 & 170  & 1.59 & 11.3 & 2.86     &  14.6\\
9 &   0.0  & 274.1 & $-$  & $-$ & $-$  & $-$  & $-$       & $-$  & 2368 & 282  & 1.55 & 10.7 & 2.53     & 12.8 & $-$  & $-$  & $-$  & $-$  & $-$      &  $-$ \\
10&   0.0  & 365.5 & $-$  & $-$ & $-$  & $-$  & $-$       & $-$  & 2849 & 317  & 1.60 & 10.8 & 2.57     & 12.7 & $-$  & $-$  & $-$  & $-$  & $-$      &  $-$ \\
\hline
11&   0.1  &-18.3  & 392  & 188 & 1.65 & 10.1 & 1.96 & 11.7 & 56 & 39& 1.86 & 11.7     & $1.86^*$ & 11.8 & $-$    & $-$    & $-$  & $-$  &  $-$ &  $-$\\
12&   0.1  &-9.1   & 450  & 213 & 1.69 & 10.1 & 2.02 & 11.6 & 146 & 79 & 1.88 & 11.9   & $1.93^*$ & 12.5 & 166  & 43   &  1.84 & 12.0 & 1.86$^*$ & 12.4\\
13&   0.1  & 0.0   & 556  & 230 & 1.64 & 9.9  & 2.06 & 11.6 & 210 & 106 & 1.88 & 11.8  & $2.05^*$ & 13.1 & 222  & 59   &  1.86 & 12.2 & 2.01 & 14.3\\
14&  0.1   & 9.1   & 677  & 245 & 1.61 & 9.7  & 2.08 & 11.5 & 306 & 127 & 1.80 & 11.4  & 2.16  & 13.1 & 289  & 72   &  1.84 & 12.1 & 2.21 & 14.5\\
15&  0.1   & 18.3  & 779  & 257 & 1.59 & 9.7  & 2.10 & 11.5 & 421 & 144  & 1.71 & 10.9  & 2.24  & 13.1 & 334  & 85   &  1.79 & 11.9 & 2.35 & 14.6\\  
16&  0.1   & 91.4  & 1383 & 331 & 1.54 & 8.7  & 2.19 & 11.4 & 1174  & 217  & 1.55 & 10.3  & 2.44  & 12.9 & 1008   & 146  &  1.58 & 11.0 & 2.77 & 14.6\\ 
17&  0.1   & 137.1 & 1537 & 367 & 1.53 & 8.8  & 2.23 & 11.3 & 1471  & 246  & 1.55 & 10.3  & 2.49  & 12.9 & 1351   & 168  &  1.59 & 11.1 & 2.86 & 14.6\\ 
18&  0.1   & 182.7 & $-$  & $-$   & $-$  & $-$  & $-$  & $-$ & 1743  & 273  & 1.57 & 10.4  & 2.52  & 12.8 & 1624   & 187  &  1.61 & 11.3 & 2.91 & 14.6\\  
19&  0.1   & 274.1 & $-$  & $-$   & $-$  & $-$  & $-$  & $-$ & 2242  & 317  & 1.60 & 10.5  & 2.57  & 12.7 & $-$    & $-$    & $-$ & $-$ & $-$ & $-$\\ 
\hline 
20&  0.2   &-18.3  & 455  & 241 & 1.76 & 10.1 & 2.08 & 11.5 & 87   & 74  & 1.98  & 12.1 & $2.00^*$ & 12.4 & 78 & 32     & 1.94 & 12.2 & 1.85$^*$ & 12.4\\
21&  0.2   &-9.1   & 593  & 259 & 1.72 & 9.9  & 2.10 & 11.5 & 174  & 113 & 1.97  & 11.9 & $2.11^*$ & 13.0 & 170   & 56  & 1.96 & 12.4 & 2.02$^*$ & 13.3\\
22& 0.2    & 0.0   & 670  & 274 & 1.70 & 9.8  & 2.13 & 11.5 & 282  & 139 & 1.89  & 11.4 & 2.22 & 13.1 & 227   & 73  & 1.95 & 12.4 & 2.21 & 14.5\\
23& 0.2    & 9.1   &  793 & 287 & 1.68 & 9.8  & 2.14 & 11.5 & 407  & 157 & 1.80  & 11.1 & 2.28 & 13.1 & 291   & 87  & 1.91 & 12.1 & 2.35 & 14.6\\
24& 0.2    & 18.3  &  850 & 299 & 1.66 & 9.7  & 2.16 & 11.4 & 528  & 172 & 1.74  & 10.8 & 2.33 & 13.1 & 380   & 100 & 1.83 & 11.8 & 2.77 & 14.6\\  
25& 0.2    & 91.4  & 1357 & 369 & 1.61 & 8.8  & 2.23 & 11.3 & 1145   & 244 & 1.64  & 10.4 & 2.48 & 12.9      & 1014    & 161 & 1.66 & 11.1 & 2.86 & 14.6\\ 
26& 0.2    & 137.1 &   $-$ & $-$  & $-$  & $-$  & $-$ & $-$ & 1438   & 276 & 1.63  & 10.4   & 2.52 & 12.8     & 1307  & 185 & 1.67 & 11.2 & 2.91 & 14.6\\ 
27& 0.2    & 182.7 &   $-$ & $-$  & $-$  & $-$  & $-$ & $-$ & 1713   & 304 & 1.64  & 10.4 & 2.55 & 12.7     &  $-$    & $-$   & $-$ & $-$ & $-$ & $-$ \\
28& 0.2    & 228.4 &   $-$ & $-$  & $-$  & $-$  & $-$ & $-$ & 1928   & 329 & 1.65  & 10.5 & 2.58 & 12.7 &  $-$    & $-$   & $-$ & $-$ & $-$ & $-$\\ 
\hline
29& 0.3    & -18.3 & 138   & 118  & 1.80 &10.0  & 2.15 & 11.5 & 158  & 103 & 2.01 & 12.2 & $2.07^*$ & 12.8 & 104  & 46  & 2.04 & 12.6 & 2.06$^*$ & 12.4\\
30& 0.3    & -9.1  &  262  & 149  & 1.78 &9.9   & 2.17 & 11.4 & 291  & 154 & 2.03 & 12.0 & 2.28 & 13.1 & 175  & 70  & 2.05 & 12.7 & 2.19 & 14.1\\
31& 0.3    & 0.0   & 699   & 319  & 1.76 &9.9   & 2.18 & 11.4 & 383  & 169 & 1.88 & 11.2 & 2.32 & 13.1 & 235  & 88  & 2.02 & 12.5 & 2.39 & 14.6\\
32& 0.3    & 9.1   & 820   & 332  & 1.75 &9.8   & 2.20 & 11.4 & 501  & 185 & 1.83 & 10.9 & 2.37 & 13.0 & 316  & 103 & 1.94 & 12.1 & 2.52 & 14.6\\
33& 0.3    & 18.3  & 896   & 345  & 1.74 &9.8   & 2.21 & 11.3 & 595  & 198 & 1.80 & 10.8 & 2.40 & 13.0 & 410  & 116 & 1.85 & 11.6 & 2.71 & 14.6\\
34& 0.3    & 91.4  & $-$   & $-$  & $-$  & $-$  & $-$  & $-$ & 1147  & 272  & 1.72 & 10.5 & 2.52 & 12.8 & 1013   & 176 & 1.74 & 11.2 & 2.88 & 14.6\\
35& 0.3    & 137.1 & $-$   & $-$  & $-$  & $-$  & $-$  & $-$ & 1411  & 306  & 1.72 & 10.5 & 2.52 & 12.7 & $-$    & $-$   & $-$ & $-$ & $-$ & $-$ \\
36& 0.3    & 182.7 & $-$   & $-$  & $-$  & $-$  & $-$  & $-$ & 1660  & 336  & 1.72 & 10.5 & 2.55 & 12.6 & $-$    & $-$   & $-$ & $-$ & $-$ & $-$   \\
\toprule 
\end{tabular}
\caption{List of the models analyzed in this work: for each hadronic EOS (soft, intermediate or stiff) we explore different values of the NJL parameters  $\delta\Omega_0$ and $g_v/g_s$. The phase transition parameters  $\Delta \varepsilon$ and $P_t$ are given in $\mathrm{MeV} ~ \mathrm{fm}^{-3}$.  The terminal mass $M_{\mathrm{T}}$  of the last stable hybrid star configuration is given  in units of $M_{\odot}$ and the corresponding terminal radius $R_{\mathrm{T}}$ in $\mathrm{km}$. The maximum mass $M_{\mathrm{max}}$  and the corresponding radius $R_{\mathrm{max}}$ are also in units of $M_{\odot}$ and $\mathrm{km}$, respectively. The asterisks indicate models for which $M_{\mathrm{max}}$ is not in a ``cusp''.}
\label{table2}
\end{table*}

\section{The equation of state}
\label{sec:2}

As mentioned before, the EOS of cold dense matter is well founded on nuclear theory and  experiments for densities $\lesssim 1 \, n_0$. Also, perturbative QCD (pQCD) can describe deconfined quark matter accurately above $\sim 40\,n_0$  \citep{Kurkela:2009gj, Gorda:2018gpy}.   Between these limits, model-informed and  model-agnostic approaches have been used in the literature to describe the EOS. In this work we will use a generalized piecewise polytropic (GPP) EOS for hadronic matter and the Nambu-Jona-Lasinio model for quark matter. Both models are briefly summarized in the following subsections.

\subsection{The hadronic phase}   
\label{sec:EOS_H}

Piecewise polytropic EOSs have been used in several works to study many properties of NSs in a quite model independent way \cite{Hebeler:2013nza,OBoyle-etal-2020,Annala:2020efq}. Here we use the generalized piecewise polytropic (GPP) model presented in Ref.  \cite{OBoyle-etal-2020}.  
For the hadronic crust we use a GPP fit to the SLy(4) EOS found in Ref.~\cite{Douchin:2001sv} which accurately reproduces the EOS and the adiabatic index.
For the hadronic part of the core with baryon number density above $0.3\, n_0$ we construct three different model-agnostic GPP EOSs using the prescription of Ref.~\cite{OBoyle-etal-2020}  to ensure continuity in the pressure, the energy density, and the  speed of sound  $c_\mathrm{s}$. \
 {The EOS parameters are chosen arbitrarily to obtain a soft, an intermediate and a stiff EOS that match at $1.1~n_0$ with predictions of  chiral effective field theory (cEFT) interactions, which has given a frame for a systematic expansion of nuclear forces at low momenta explaining the hierarchy of two-, three-, and weaker higher-body forces \cite{Coraggio:2012ca,Gandolfi:2011xu,Holt:2012yv,Hebeler:2009iv,Sammarruca:2012vb,Tews:2012fj}. At present, microscopic calculations based on cEFT interactions allow a reliable determination of the properties of neutron star matter up to the nuclear saturation density $n_0$;  specifically,  the pressure is currently known to roughly $\pm 20\%$ accuracy at $n_0$ \cite{Tews:2012fj,Hebeler:2009iv}. In the regime of intermediate densities above $n_0$, our understanding of the EOS is insufficient, but strong constraints can be imposed from $2 \, M_{\odot}$ pulsars, causality and thermodynamic consistency.}

\begin{figure*}[tb]
\centering
\includegraphics[scale=0.50]{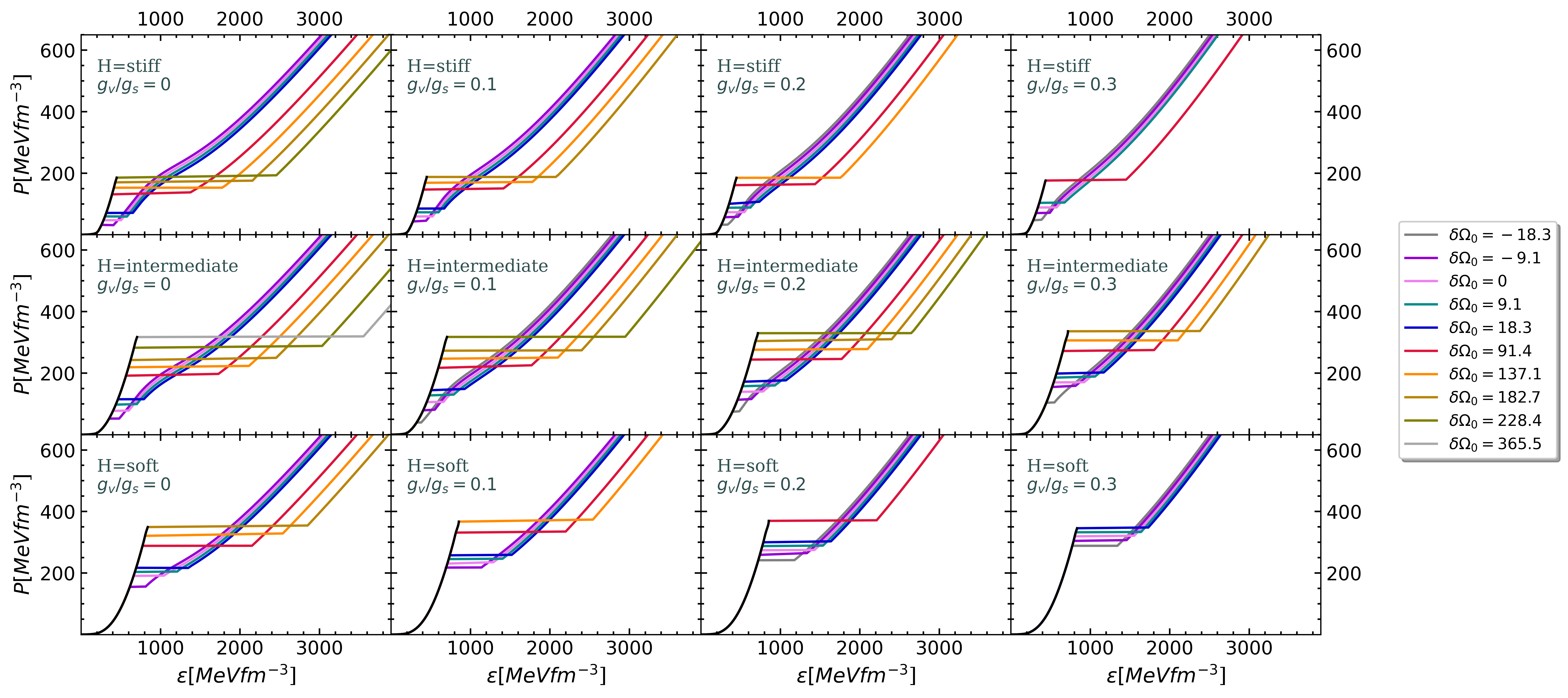}
\caption{Hybrid EOSs used in this work. For each of the three hadronic EOS labeled as soft, intermediate and stiff we consider different phase transitions using different values of the NJL parameters $\delta \Omega_0$ and $g_v$.   The value of $\delta \Omega_0$ (in $\mathrm{MeV ~fm^{-3}}$) is indicated by the color codes shown on the right. Within each panel we indicate the hadronic model and the value of $g_v$.}
\label{fig:EOS}
\end{figure*}

\subsection{The quark phase}  
\label{sec:EOS_Q}

For the quark matter phase we use a SU(3) NJL model with scalar-pseudoscalar, isoscalar-vector and  't Hooft six fermion interactions. The Lagrangian density of the model is \cite{Buballa:2003qv,Masuda:2012ed}:
\begin{equation}
\begin{aligned}
{\cal L}  = & \bar\psi(i \gamma_\mu \partial^\mu - \hat{m})\psi \\ 
& +  g_s \sum_{a=0}^{8} [( \bar\psi \lambda^a \psi )^2  +  (\bar \psi i \gamma_5 \lambda^a \psi)^2 ] \\
& -   g_v \sum_{a=0}^{8}   [(\bar\psi \gamma_\mu  \lambda^a \psi)^2 +  (\bar\psi \gamma_5 \gamma_\mu \lambda^a\ \psi)^2 ]  \\
& +   g_t\{\det[ \bar\psi(1+\gamma_5)\psi]   +  \det[ \bar\psi(1-\gamma_5)\psi]\},  
\label{eq2}
\end{aligned}
\end{equation}
where $\psi = (u,d,s)$ denotes the quark fields,  $\lambda^a ( 0 \leq a \leq 8 )$ are  the $U(3$) flavor matrices,  $\hat{m} = \mathrm{diag}({m}_{u},{m}_{d},{m}_{s})$ is the quark current mass, and $g_s$, $g_v$ and $g_t$ are coupling constants.  We have not included diquark interactions in the Lagrangian because we focus in the regime of weak diquark coupling where color superconductivity is shifted to very large densities that do not occur at NS cores. 

The mean-field thermodynamic potential density $\Omega$ for a given baryon chemical potential $\mu$ at $T = 0$, is given by
\begin{eqnarray}
\Omega  = & - & 6  \sum_i\int_{k_{Fi}}^{\Lambda}{\frac{p^2\, dp}{2\pi^2}}\sqrt{p^2+M_i^2} + 2g_s\sum_i \langle \bar \psi \psi \rangle_i^2  \nonumber \\
   & - & 2 g_v \sum_i \langle \psi^\dagger \psi \rangle_i^2 + 4g_t\langle\bar u u \rangle \langle\bar d d \rangle \langle\bar s s \rangle \nonumber  \\ 
  &  - & 6 \sum_i \mu_{i}{\int_0^{k_{Fi}}  \frac{p^2\, dp}{2\pi^2}}-\Omega_0, 
\label{eq3}
\end{eqnarray}
where the sums run over the quark flavor $(i = u, d, s)$ and  $\Lambda$ is a regularization ultraviolet cutoff introduced to avoid divergences in the medium integrals. The Fermi momentum of the  $i$-species is given by
$$k_{Fi} = \theta(\mu^{\ast}_{i} - M_i)\sqrt{ \mu^{\ast 2}_{i} - M_i^2},$$ 
where $\mu^\ast_{i}$ is the quark chemical potential modified by the vector interaction: 
$$\mu^{\ast}_{u,d,s} = \mu_{u,d,s} - 4 g_v \langle \psi^\dagger \psi \rangle_{u,d,s}.$$
The constant $\Omega_0$ is usually fixed by imposing the condition that $\Omega(T=0, \mu=0) =0$ \cite{Buballa:2003qv}.

For the EOS parameters, we adopt  $\Lambda = 631.4 \, \mathrm{MeV}$, $g_s\Lambda^2 = 1.829$, $g_t\Lambda^5 = -9.4$, $m_u = m_d = 5.6 \, \mathrm{MeV}$, $m_s = 135.6 \, \mathrm{MeV}$, in order to fit the vacuum values of the pion mass, the pion decay constant, the kaon mass, the kaon decay constant, and the quark condensates \citep{HATSUDA:1994,Kunihiro:1989my,Ruivo:1999pr}. The vector coupling constant $g_v$ is treated a free parameter.  Matter is assumed to be charge neutral and in chemical equilibrium under weak interactions. 

An important issue is the ansatz adopted for fixing $\Omega_0$ in Eq. (\ref{eq3}). The standard assumption that $\Omega$ must vanish at $\mu = T =0$  leads to the value $\Omega_0 = 5076 \, \mathrm{MeV \, fm}^{-3}$ for the above quoted parametrization. 
However, despite being reasonable, this prescription is arbitrary and other choices are possible \citep{Schertler:1999xn, Pagliara:2007ph, Lenzi:2012xz} (notice that in the MIT bag model for instance, the pressure at $\mu = T =0$ is non-vanishing). 
In view of this,  a different procedure was adopted in Ref. \cite{Pagliara:2007ph}:  the parameter  $\Omega_0$ was chosen in such a way that  the hadron-quark deconfinement  occurs at the same chemical potential as the chiral phase transition. Another approach, is to regard $\Omega_0$ as a free parameter and explore the consequences of varying it \cite{Lenzi:2012xz}. Following Ref. \cite{Lenzi:2012xz}, we will replace $\Omega_0$ in Eq. (\ref{eq3}) by the new value $\Omega_0 + \delta \Omega_0$, being $\Omega_0 = 5076 \, \mathrm{MeV \, fm}^{-3}$ and taking $\delta \Omega_0$ as a free parameter. Notice that the chemical potential at which the chiral transition occurs is determined from the solution of the gap equations for the constituent masses and therefore, it doesn't depend on the value of $\delta \Omega_0$.

\subsection{Hybrid equations of state}

We will construct hybrid EOSs assuming that the interface between hadrons and quarks is a sharp discontinuity described by the Maxwell construction; i.e. the pressure and the Gibbs free energy per baryon are the same at both sides of the phase splitting surface.  
To this end we combine the three hadronic models described in Sec. \ref{sec:EOS_H} with different parametrizations of the NJL model of Sec. \ref{sec:EOS_Q}. 

 {As mentioned in the previous subsection, we used the HK parameter set \citep{HATSUDA:1994,Kunihiro:1989my,Ruivo:1999pr}  to fix the quark and pseudoscalar meson properties in the three-flavor NJL model. It is worth noting that alternative parameterizations, such as the RKH \cite{Rehberg:1996} or LKW \cite{LUTZ:1992}, could have been selected. However, to minimize the number of free parameters, we chose to vary only the ``bag constant'' $\delta \Omega_0$ and the vector coupling constant $g_v$ in this study. This choice is justified for two reasons: Firstly, vector meson masses are less accurately determined than pseudoscalar meson masses \cite{Vogl:1991qt}. Secondly, varying the parameters of pseudoscalar mesons has a lesser impact on the EOS than varying $g_v$. For instance, after fixing the hadronic EOS, the critical baryon chemical potential and energy density jump $\Delta \epsilon \equiv \epsilon_{q} - \epsilon_{h}$ at the quark-hadron interface are nearly identical for HK, RKH, and LKW parameterizations (see Table III of \cite{Khanmohamadi:2019jky}). 
On the contrary, $\delta \Omega_0$ and $g_v$ have a strong influence on the transition pressure $P_t$ and the energy density jump $\Delta \epsilon$, as we will see below.
}

From previous works \cite{Pagliara:2007ph, Lenzi:2012xz}, we know that  $\delta \Omega_0$ has a minimum value because the phase transition cannot be shifted to a pressure regime where the NJL model describes the vacuum (this value is $\delta \Omega_0 = - 18.3 \, \mathrm{MeV} \, \mathrm{fm}^{-3}$, as shown in Table \ref{table2}). 
On the other hand, a maximum value of $\delta \Omega_0$ results from the fact that the hadronic and quark curves do not intersect if $\delta \Omega_0$ is too large. For each hadronic parametrization and each value of the vector coupling constant $g_v$,  we explored $\delta \Omega_0$ up to the largest possible value.

The value of the vector coupling constant $g_v$ is still uncertain, although efforts have been made to determine it in various ways. Values in the range $0.25<g_v/g_s<0.5$ were derived by a Fierz transformation of effective one-gluon exchange interaction, with $g_{v}$ depending on the strength of the $U_{A}(1)$ anomaly in the two-flavor model \cite{Klevansky:1992qe,Kashiwa:2011td}.  Other attempts to estimate $g_{v}$ are based on the fitting of the vector meson spectrum \cite{Vogl:1991qt}. However, as discussed in Refs. \cite{Fukushima:2008wg, Zhang:2009mk}, the relation between the vector coupling in dense quark matter and the meson spectrum in vacuum is expected to be strongly modified by in-medium effects. 
In summary, there is agreement that $g_{v} / g_{s}$ should be of the order of unity but there is no consensus on the precise value of $g_v$. In this work we adopted $g_v / g_s =  0, 0.1, 0.2, 0.3$ because for larger values the quark EOS gets too stiff preventing in most cases an intersection between the Gibbs free energy per baryon of the hadronic and quark phases (a similar situation was found in Ref.  \cite{Lenzi:2012xz}).

The curves of the resulting hybrid EOSs are shown in Fig. \ref{fig:EOS}. As seen in Table \ref{table2}, the transition pressure $P_t$ ranges from low values around $\sim 30 ~\mathrm{MeV ~fm^{-3}}$ to high values above $\sim 350 ~\mathrm{MeV ~fm^{-3}}$.  The energy density jump at the interface $\Delta \epsilon \equiv \epsilon_Q(P_t) - \epsilon_H(P_t)$ varies from  $\sim 50 ~\mathrm{MeV ~fm^{-3}}$ to   $\sim 1700 ~\mathrm{MeV ~fm^{-3}}$. The exact values for each model are shown in Table \ref{table2}.

\begin{figure*}[tb]
\centering
\includegraphics[scale=0.50]{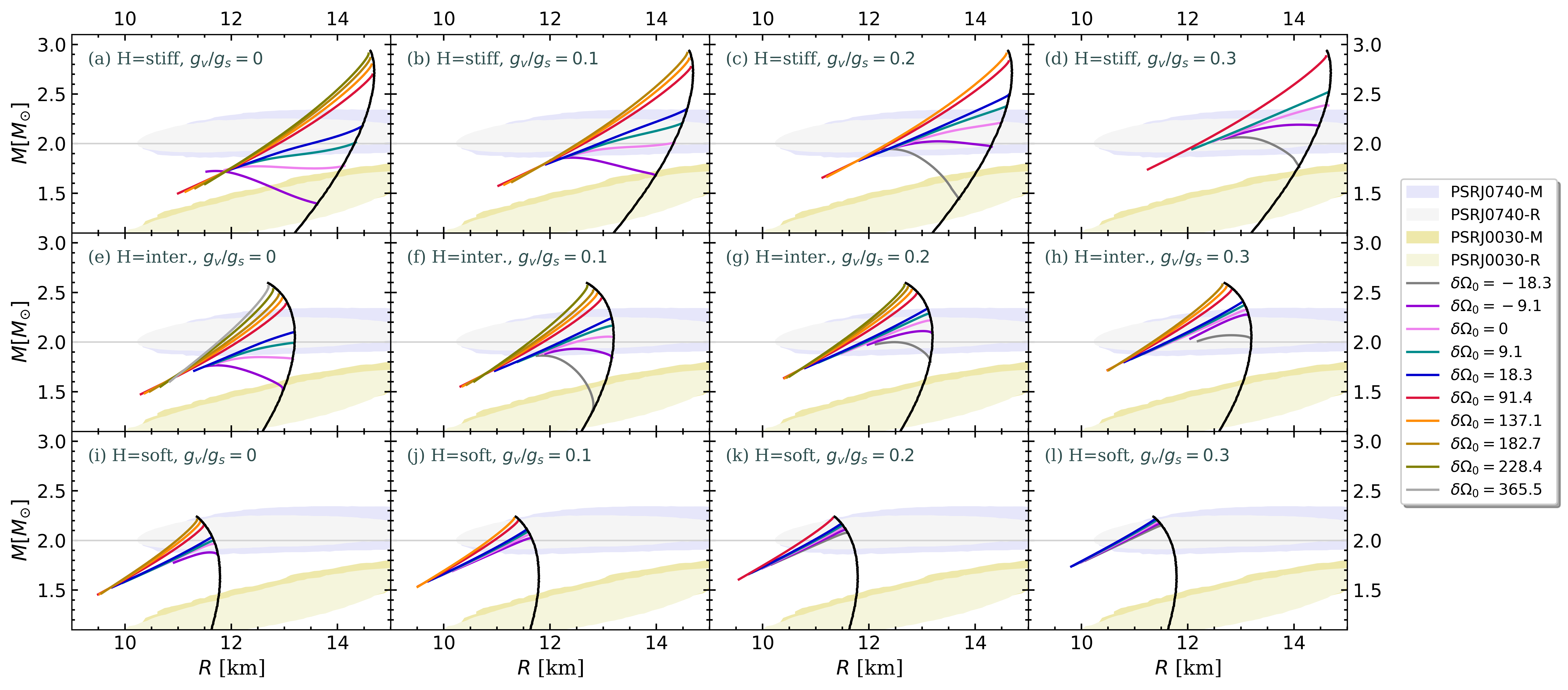}
\caption{Mass-radius relations for the hybrid EOSs shown in Fig. \ref{fig:EOS} (same color-coding is used). We also show astrophysical constraints from the $\sim 2 \, M_\odot$ pulsars and NICER \citep{Riley:2019yda,Miller:2019cac,Riley:2021pdl,Miller:2021qha} observations. When slow interface conversions are assumed all points of the  curves represent dynamically stable stars. If interface conversions are rapid, hybrid segments to the left of the maximum mass are unstable (see text).  {We retained some models that do not produce pulsars  above $\sim 2 \, M_\odot$ to illustrate the need for precise parameter tuning to achieve traditional hybrid stars that are consistent with such observations.}  }
\label{fig:MR}
\end{figure*}

\section{The mass-radius relation of hybrid stars with reactive interfaces}
\label{sec:MR}

To construct hydrostatic stellar configurations, we solve the Tolman–Oppenheimer–Volkoff equations. For each configuration we determine whether it is dynamically stable; i.e.  we check if the frequency $\omega_{0}$ of the fundamental radial oscillation mode verifies $\omega_{0}^{2} > 0$. For compact stars composed of cold catalyzed matter it has been shown that changes of dynamic stability occur at maxima or minima in the $M-\epsilon_{c}$ diagram, being $M$ the stellar mass and $\epsilon_{c}$ the energy density at the stellar center \cite{Harrison1965}. Thus, the analysis of the sign of $\omega_{0}^2$  can be replaced by the much more simple static analysis of the derivative $\partial M/\partial \epsilon_{c}$. However, in the case of hybrid stars with reactive discontinuities, the stability analysis must take into account that  radial perturbations may induce the phase conversion of fluid elements in the neighborhood of the interface. In such a case, it may happen that matter becomes non-catalyzed, and dynamic stability cannot be assessed by the derivative $\partial M/\partial \epsilon_{c}$. As discussed in Refs. \cite{Haensel1989,Karlovini:2003xi,Pereira:2017rmp}, there are two essentially different behaviors depending on the reaction timescale $\tau$ at the interface. If $\tau$ is much shorter than the oscillation period (rapid conversions), fluid elements change  instantaneously and periodically from one phase to the other as the pressure oscillates around $P_t$.  However, if $\tau$ is much larger than the oscillation period (slow conversions), the motion around the interface involves only the stretch and squash of volume elements without any phase transformation. Since the period of the fundamental radial oscillation mode of neutron stars is always around $\sim 1 \mathrm{ms}$  \cite{Gondek:1997fd, Benvenuto:1998tx, VasquezFlores:2010eq, VasquezFlores:2012vf, Brillante:2014lwa, Flores:2016eyg, DiClemente2020, Jimenez:2021wil}, oscillations would be rapid or slow depending on whether the conversion timescale is respectively much smaller or much larger than $\sim 1 \mathrm{ms}$.

The first analysis of the effect of the hadron-quark conversion speed on the dynamic stability of hybrid stars was carried out in Ref. \cite{Pereira:2017rmp} focusing on the limiting cases of rapid and slow conversions. In these two cases, all the complexity of the phase changing mechanism can be encoded into simple junction conditions on the radial  fluid displacement $\xi$ and  the corresponding Lagrangian perturbation of the pressure $\Delta p$ at the interface  (see \citep{Haensel1989, Karlovini:2003xi, Pereira:2017rmp} for further details on the junction conditions). As shown in \citep{Pereira:2017rmp}, changes of stellar stability still occur only at maxima or minima in the $M-\epsilon_{c}$ diagram if phase conversions are rapid. However, in the slow conversion scenario, changes of stability do not occur necessarily at critical points. In fact, it turns out that $\omega_{0}$ can be a real number (indicating stability) even if $\partial M / \partial \epsilon_{c}<0$ \cite{Pereira:2017rmp}. Thus,  many configurations that were believed to be radially unstable are in fact stable under small perturbations in the slow conversion scenario. These configurations have been called slow-stable hybrid stars (SSHSs) \cite{Lugones:2021bkm}. The star for which $\omega_{0}^2$ vanishes is the last stable object of the SSHS branch and will be called \textit{terminal configuration} (with a terminal mass $M_T$ and a terminal radius $R_T$.)

In Fig.~\ref{fig:MR} we show our results for the $M$-$R$ relation using the hybrid EOSs presented in Fig. \ref{fig:EOS}. All points of the curves represent dynamically stable configurations if hadron-quark conversions at the interface are slow. However, if interface conversions are rapid all configurations beyond the maximum mass are unstable. 
We also include constraints coming from recent observations. The gray horizontal line takes into account the observed masses of the pulsars PSR J1614-2230 \cite{Demorest:2010bx}, PSR J0348+0432 \cite{Antoniadis:2013pzd} and PSR J0740+6620  \cite{Cromartie:2019kug}, which require that an acceptable EOS must be able to support a NS of at least $2 \, M_{\odot}$.
Additionally,  we show the  $95\%$ confidence intervals for the mass and radius of the millisecond-pulsars PSR J0030+0451 \cite{Riley:2019yda,Miller:2019cac} and PSR J0740+6620 \cite{Riley:2021pdl,Miller:2021qha} measured recently by the Neutron Star Interior Composition Explorer (NICER).

\begin{figure*}[tb]
\centering
\includegraphics[scale=0.50]{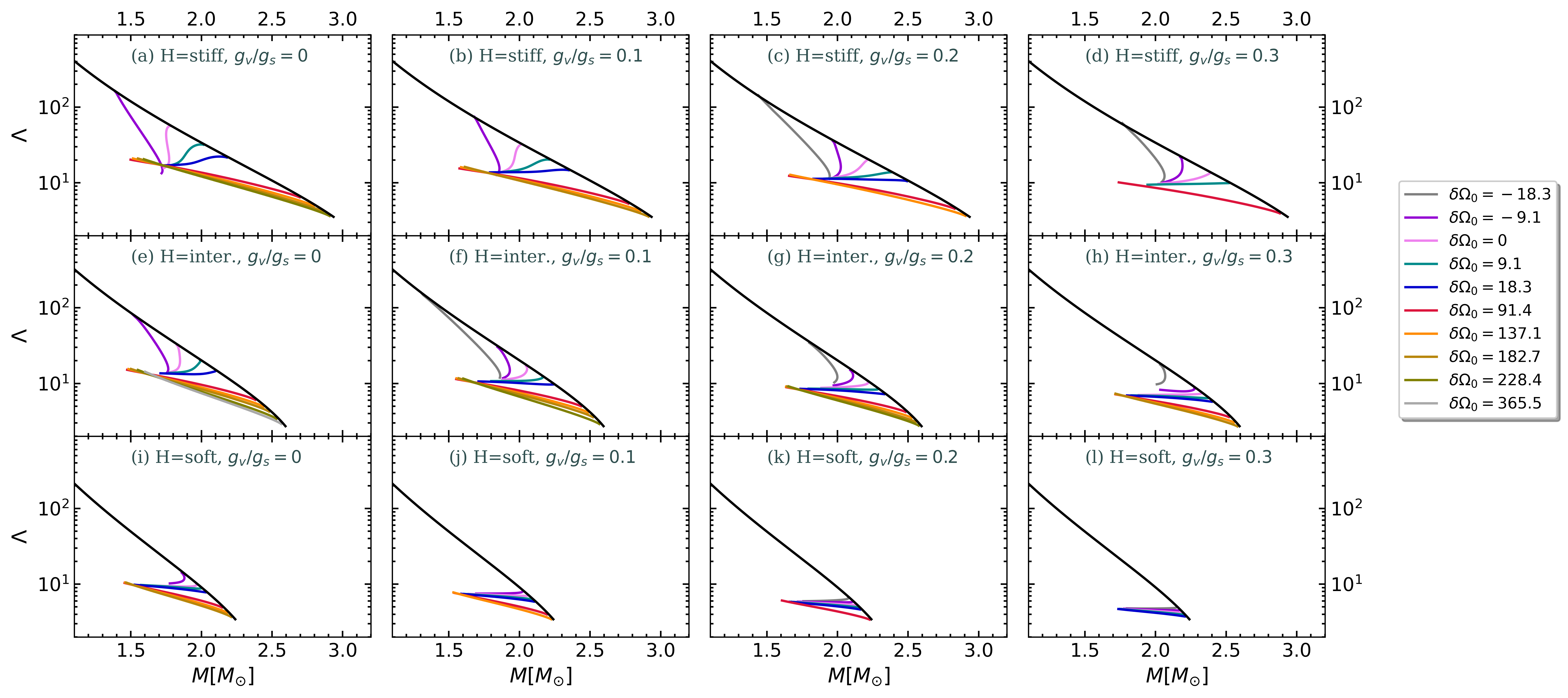}
\caption{Dimensionless tidal deformability $\Lambda$ as a function of the gravitational mass $M$ for the hybrid EOSs of Fig. \ref{fig:EOS} (same color-coding). 
For slow conversions, all points of the curves represent stable stars and $\Lambda$ can be either an increasing or a decreasing function of $M$. For rapid conversions, models beyond the maximum mass are unstable and $\Lambda$ is always a decreasing function of $M$.  }
\label{fig:LM}
\end{figure*}

In most curves of Fig.~\ref{fig:MR} there is a cusp at the maximum mass point where the pure hadronic branch and the SSHS branch coincide.  In these cases, the transition pressure is high, and obviously the higher the $P_t$, the greater the maximum mass of the curve. A long SSHS branch exists to the left of the cusp and  consequently twin stars with the same $M$ but different radii are possible for a wide range of masses. As seen in Fig. \ref{fig:MR} and Table \ref{table2}, in most cases the larger the $P_t$ and the $\Delta \varepsilon$, the larger the length of the SSHS branch. 
In the scenario with a mass cusp, the hadronic and the SSHSs  branches fulfill NICER and $2 \, M_{\odot}$ pulsar constraints for a broad set of values of the EOS parameters, as seen in Fig. \ref{fig:MR}.

On the other hand, there are some curves (without a cusp) where the maximum mass of the sequence is fully in the hybrid sector. However, notice that it is difficult to fulfill the $2 \, M_{\odot}$ constraint in these cases. 
This scenario occurs for large enough values of the vector coupling constant ($g_v/g_s = 0.2, 0.3$)  and small values of the bag constant ($\delta \Omega_0 = -18.3, -9.1, 0$) as seen in panels (c), (d), (g) and (h) of Fig.~\ref{fig:MR}. In Table \ref{table2}, models of this class are identified with an asterisk. 
The relatively small value of $\delta \Omega_0$ decreases the value of $P_t$ with respect to models of the same panel (i.e. keeping all other parameters fixed). As a consequence, hybrid configurations are possible at smaller stellar masses. Concomitantly, a large enough value of $g_v$ makes the quark EOS stiff enough to allow a maximum mass above $2 \, M_{\odot}$. 
Within this kind of models we find hybrid stars that are \emph{totally stable}, i.e. they are dynamically stable in both the rapid and the slow conversion scenarios. These configurations are located to the right of the maximum mass object and fulfill the condition $\partial M / \partial \epsilon_c > 0$.
To the left of the maximum mass, we find hybrid star configurations that are dynamically stable only in the slow conversion scenario. The length of these slow-stable branches is in general much shorter than in the models with cusps, i.e. the difference  $\Delta M \equiv M_{\mathrm{max}} - M_{\mathrm{T}}$ is relatively small.  Also in this case, $\Delta M$ increases with the increase of $P_t$ and $\Delta \varepsilon$.

\section{Tidal deformability of SSHS}
\label{sec:tidal}

Neutron stars are tidally deformed during the inspiral phase of a NS-NS merger, leaving a detectable imprint on the observed gravitational waveform of the event. This effect can be quantified in terms of the so-called dimensionless tidal deformability of the star, defined as: 
\begin{align}
\Lambda_{i}=\frac{2}{3} k_{2}^{(i)}\left(\frac{c^{2} R_{i}}{G M_{i}}\right)^{5},
\end{align}
being $k_{2}^{(i)}$ the second Love number, $R_{i}$ the radius, and $M_{i}$ the mass of the $i$-th star \cite{Flanagan:2007ix}. The tidal deformability $\Lambda_{i}$ describes the amount of induced mass quadrupole moment when reacting to a certain external tidal field \cite{Hinderer:2007mb,Damour:2009vw}.
 {To perform a theoretical calculation of the tidal deformability, we employ the methodology outlined in Ref. \cite{Parisi:2020qfs}, which involves an accurate treatment of the finite energy-density discontinuity \cite{Postnikov:2010yn} and incorporates more recent corrections to the junction condition \cite{Takatsy:2020bnx, Zhang:2020pfh}.}

 {The primary constraint on the value of $\Lambda$ currently comes from the detection of gravitational waves resulting from the NS-NS merger event GW170817 by LIGO/Virgo \cite{TheLIGOScientific:2017qsa, Abbott:2018exr, Abbott:2018wiz}. Assuming both objects have spins within the observed range in Galactic binary NSs, different limits can be obtained for the dimensionless tidal deformability, depending on additional hypotheses. Ref. \cite{Abbott:2018exr} made assumptions about the behavior of the EOS, such as assuming the same EOS for both NSs and using EOS-independent universal relations, which may not hold for SSHSs. Therefore, we do not compare our calculations with that constraint in this work. Instead, we consider the constraints derived in the discovery paper of GW170817 \cite{TheLIGOScientific:2017qsa} and the subsequent updated analysis \cite{Abbott:2018wiz}. These works inferred the binary parameters from the inspiral signal, making minimal assumptions about the nature of the compact objects. Specifically, the dimensionless parameters $\Lambda_i$ governing the tidal deformability of each component were given a prior distribution uniform within $0 \leq \Lambda_i \leq 5000$, with no assumed correlation between $\Lambda_1$, $\Lambda_2$, and the mass parameters \cite{Abbott:2018wiz}. This approach implicitly allows each neutron star to have a different EOS and includes the possibility of phase transitions within the stars, exotic compact objects, or even black holes as binary components \cite{Abbott:2018wiz}. Their limits for GW170817 are therefore appropriate for comparison with the SSHS results presented in this work.}

In Fig.~\ref{fig:LM}  we show $\Lambda$ as a function of $M$ for the hybrid EOSs of Fig.~\ref{fig:EOS}.  For hadronic stars, the larger the mass the smaller the $\Lambda$ \citep{Chatziioannou:2018vzf}.  However, for hybrid stars with slow hadron-quark conversions at the interface, $\Lambda$ can decrease or increase with $M$ meaning that, for a given hybrid EOS, not necessarily the most massive component of a binary NS merger will have the smallest $\Lambda$. If interface conversions are rapid,  models beyond the maximum mass are unstable and $\Lambda$ has the standard behavior, i.e. it decreases with $M$.

\begin{figure}[tb]
\centering
\includegraphics[width=\columnwidth]{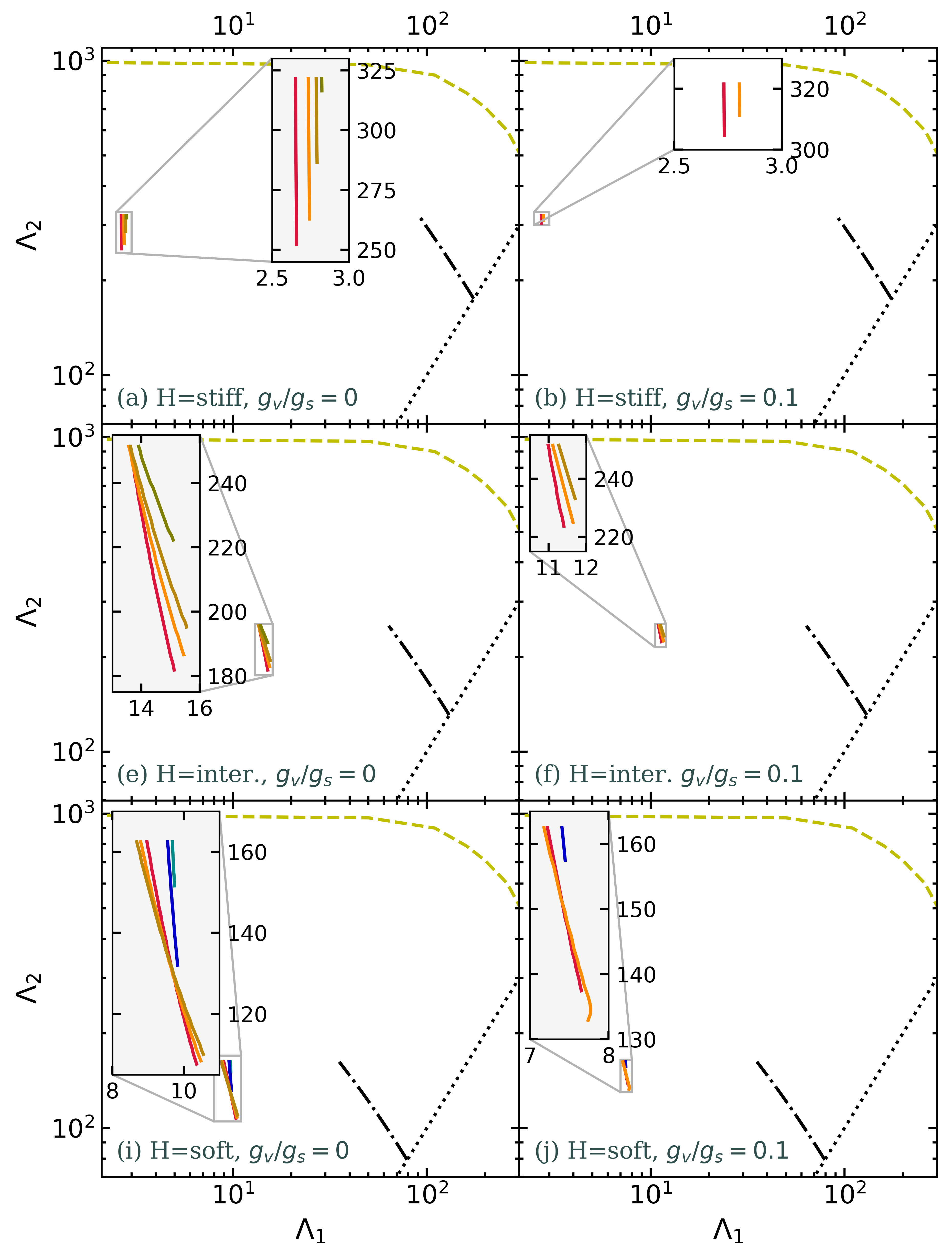}
\caption{Dimensionless tidal deformabilities $\Lambda_{1}$ and $\Lambda_{2}$ for a binary NS system with masses $M_1$ and $M_2$ ($M_1 > M_2$) and the same chirp mass and mass ratio as GW170817 \cite{Abbott:2018wiz}. In all cases the maximum mass of the model is $> 2 \, M_{\odot}$. The diagonal dotted line indicates the $\Lambda_{1}=\Lambda_{2}$ boundary and the dashed line denotes the $50 \%$ level determined by LIGO/Virgo in the low-spin prior scenario. Black dash-dotted lines represent mergers with two hadronic stars. Colored solid lines indicate the merging of a SSHS of mass $M_1$ and a hadronic star with mass $M_2$ (color codes have the same meaning as in the previous figures).   {The merger of two SSHS is not compatible with GW180718 due to the absence of SSHSs in the secondary object's mass range $1.16 < M_2 [M_{\odot}] < 1.36$ (cf. Fig. \ref{fig:MR})}. }
\label{fig:L1L2}
\end{figure}

In Fig.~\ref{fig:L1L2}, we show the $\Lambda_1$-$\Lambda_2$ relation for  binary neutron star mergers with the same chirp mass and the same mass ratio  of  the GW170817 event   ($M_C = (M_1 M_2)^{3/5} / (M_1+M_2)^{1 / 5} = 1.188 \, M_{\odot}$ and $q = M_2/M_1$ in the range $0.7-1$).  
The $50\%$ credible region of GW170817 always contains mergers with two purely hadronic stars. However, as the hadron-hadron case has been extensively studied in the literature, it will not be further discussed here.
Focusing only on EOSs that result in $M_{\mathrm{max}}> 2~M_{\odot}$, we find that mergers of a hadronic star and a SSHS fall inside the $50\%$ region of GW170817 for several EOS parametrizations.  This scenario is possible provided that the vector coupling constant of the NJL model is sufficiently small ($g_v/g_s = 0, 0.1$). 
 {For larger values of $g_v/g_s$  (e.g. $0.2, 0.3$) the masses of SSHSs are always above $1.6 ~ M_{\odot}$ (cf. Fig. \ref{fig:MR}), i.e. out of the ranges imposed by $M_C$ and $q$ of GW170817 ($1.36 < M_1 [M_{\odot}] < 1.60$ and $1.16 < M_2 [M_{\odot}] < 1.36$ \cite{Abbott:2018wiz}). 
Finally, note that our analysis does not find any mergers involving two SSHSs if we restrict to models with $M_{\mathrm{max}}> 2~M_{\odot}$. This is due to the absence of SSHSs in the secondary object's mass range $1.16 < M_2 [M_{\odot}] < 1.36$ \cite{Abbott:2018wiz}, as shown in Fig. \ref{fig:MR}. }

\bigskip

\section{Summary and conclusions}
\label{sec:4}

In this work we studied hybrid star models with reactive quark-hadron interfaces and confronted them with current observational constraints. 
Several aspects of hybrid stars with sharp discontinuities have been studied in recent years (see e.g. Refs. \cite{Alford:2004pf, Negreiros:2010tf, Alford:2013aca, Lenzi:2012xz, Burgio:2015zka, Kaltenborn:2017hus, Paschalidis:2017qmb, Li:2021sxb, Li:2019fqe, Cierniak:2020eyh, Otto:2019zjy} and references therein). In these works, although not explicitly mentioned, the conversion timescale at the interface is assumed to be \textit{rapid}, since the derivative $\partial M / \partial \epsilon_c$ is used to assess dynamic stability. 
On the other hand, the existence of  dynamically stable hybrid stars with $\partial M / \partial \epsilon_c  < 0$, associated with \textit{slow} quark-hadron interface conversions, was first reported 
very recently in Ref.  \cite{Pereira:2017rmp}.  Because of this reason, SSHSs have been, comparatively, much less studied  (see \cite{Parisi:2020qfs, DiClemente2020, Tonetto:2020bie, Lugones:2021bkm, Mariani:2019vve, Rodriguez:2020fhf, Ranea:2022bou}).

Although we presented results for rapid as well as for slow interface conversions, the main focus of the present work was on the slow scenario, which has not yet been studied using the NJL model.
We constructed a wide set of hybrid EOSs using three different model-agnostic GPP fits for hadronic matter and a NJL model with scalar, vector and ’t Hooft interactions for quark matter. For the hadronic EOS we used three different parameterizations representing stiff, intermediate and soft models. Within the quark EOS,  the vector coupling constant $g_v$ was taken as a free parameter and we assumed (as in Ref. \cite{Lenzi:2012xz}) that there is an arbitrary split between the deconfinement and the chiral phase transitions that can be tuned by a proper choice of the bag constant ($\delta \Omega_0$). For $g_v/g_s$ we used values between $0$ and $0.3$ because for larger values the quark EOS gets too stiff making a transition between hadron and quark matter impossible.

Our study of the mass-radius relation shows that there are two types of results: ($a$)  curves with a cusp at the maximum mass point where the pure hadronic branch and the SSHS branch coincide, and ($b$)  curves without a cusp where the maximum mass of the sequence is fully in the hybrid sector (see Fig.~\ref{fig:MR}). 
In the case of type ($a$), a long SSHS branch exists to the left of the cusp and  consequently twin stars with the same $M$ but different radii are possible for a wide range of masses. The NICER and $2 \, M_{\odot}$ pulsar constraints are fulfilled for a broad set of values of the EOS parameters.
Type ($b$) curves are not found so easily because it is more difficult to fulfill the $2 \, M_{\odot}$ constraint. In this case hybrid configurations to the right of the maximum mass are dynamically stable in both the rapid and the slow conversion scenarios and fulfill the condition $\partial M / \partial \epsilon_c > 0$. Hybrid stars to the left of the $M_{\mathrm{max}}$ point are dynamically stable only in the slow conversion scenario. 
In both cases ($a$ and $b$),  the length of the slow-stable branch grows with the increase of the transition density $P_t$ and the energy density jump $\Delta \varepsilon$ at the hadron-quark interface. Our results for the $M-R$ relation  are qualitatively in agreement with those found using other EOS (cf. \cite{Pereira:2017rmp, Parisi:2020qfs, DiClemente2020, Tonetto:2020bie, Lugones:2021bkm, Mariani:2019vve, Rodriguez:2020fhf, Ranea:2022bou}).

Finally, we calculated the tidal deformability $\Lambda$ of SSHSs. We found that the behavior of $\Lambda$ as a function of $M$ is essentially the same as found in recent works on SSHSs \cite{Pereira:2017rmp,Parisi:2020qfs,DiClemente2020,Tonetto:2020bie,Lugones:2021bkm,Mariani:2019vve,Rodriguez:2020fhf,Ranea:2022bou}; i.e., unlike for hadronic stars, $\Lambda$ is not necessarily a decreasing function of $M$ (see Fig.~\ref{fig:LM}).  
 {Regarding the composition of the binary components, our study yields both similarities and differences with respect to previous research on SSHSs. In agreement with Refs. \cite{Lugones:2021bkm} and \cite{Parisi:2020qfs},  the present model can explain the GW170817 event as the result of the coalescence of either two hadronic stars or a SSHS with a hadronic star. However, contrary to the findings of model-agnostic approaches for the quark matter EOS \cite{Lugones:2021bkm}, we do not observe mergers involving two SSHSs. This is because the NJL model prohibits the existence of SSHSs in the secondary's mass range $1.16 < M_2 [M_{\odot}] < 1.36$.}

\subsection*{Acknowledgements}
G. Lugones acknowledges the financial support of the Brazilian agencies CNPq (grant 316844/2021-7) and FAPESP (grants 2022/02341-9 and 2013/10559-5).
C. H. Lenzi is thankful to the Fundação de Amparo à Pesquisa do Estado de São Paulo (FAPESP) under thematic project 2017/05660-0, Grant No. 2020/05238-9 and Power Data Tecnologia Ltda for providing a technological environment for data processing.
C. Vasquez acknowledges the Universidade Estadual da Região Tocantina do Maranhão for support. 
We are grateful for the insightful discussions with A. G. Grunfeld.

\bibliography{hybrid_NJL}

\end{document}